\documentclass[a4paper]{jpconf}
\usepackage{graphicx}
\usepackage{color}
\begin{document}
\title{Mass Enhancement and Reentrant Ground State in Magnetic Field}

\author{A. Miyake$^{1, 2}$, D. Aoki$^1$, G. Knebel$^1$, V. Taufour$^1$, J. Flouquet$^1$}

\address{$^1$ Commissariat \`a l' \'Energie Atomique, INAC, SPSMS, 17 rue des Martyrs, 38054 Grenoble, France\\
$^2$ KYOKUGEN, Osaka University, Osaka, Japan}

\ead{amiyake@cqst.osaka-u.ac.jp}	

\begin{abstract}

Three different cases of magnetic field reentrant ground states in heavy fermion systems, URhGe, UGe$_2$ and CeRhIn$_5$ recently studied in Grenoble, are discussed.
URhGe is a ferromagnetic superconductor with reentrace of superconductivity under magnetic field ($H$) which is associated to the spin reorientation field.
UGe$_2$ is a ferromagnetic superconductor with an enhancement of the upper critical field $H_{c2}$ of superconductivity at its metamagnetic transition between 2 ferromagnetic phases. 
CeRhIn$_5$ is a superconductor with $H$ reentrant antiferromagnetism.
We analyze the links between the $H$ enhancement of the different contributions to the effective mass and the field reentrant phase.
  
\end{abstract}


The unconventional superconductivity (SC) of heavy fermion compounds often occurs close to quantum singularities driven by magnetic or valence instabilities. Here we will present three cases of field reentrant phenomena linked to the interplay between field changes of their magnetic and superconducting properties.

\begin{figure}[h]
\begin{center}
\includegraphics[width=40pc]{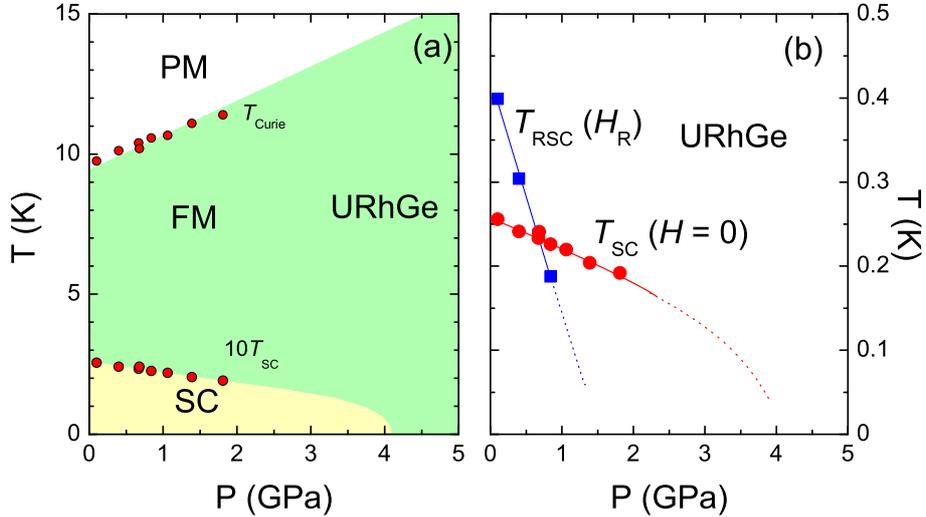}
\caption{\label{URhGe}(a) ($P, T$) phase diagram of URhGe at $H = 0$~T. (b) Pressure dependence of $T_{\rm sc}$ at $H = 0$~T (circle) and $H_{\rm R}$ (square).}  
\end{center}
\end{figure}

URhGe is a very nice example of ferromagnetic (FM) superconductor; i.e. i) its critical superconducting transition temperature $T_{\rm sc} \sim$0.27~K is far below its Curie temperature $T_{\rm Curie}\sim$ 9.5~K \cite{Aoki2001}, ii) applying a pressure drives the system deeper in the FM domain as shown in Fig.~\ref{URhGe}(a) \cite{Hardy2005, miyake2009}; thus, one escapes from a FM quantum singularity where $T_{\rm sc}\sim T_{\rm Curie}$; iii) at $T_{\rm Curie}$, the Fermi surface is well formed and thus the Fermi liquid regime as $AT^2$ resistivity law is also well established; iv) furthermore, the rather weak magnetic anisotropy leads to the remarked phenomena that the sublattice magnetization $M_0$ changes its alignment from the initial $c$-axis to the $b$-axis at $H_{\rm R} \sim$12~T along one of the hard axis (the $b$-axis) without any drastic change of $M_0\sim 0.4~\mu_{\rm B}$ \cite{Levy2005}.
The surprising result is that when $H$ increases along the $b$-axis, the reentrant superconductivity (RSC) is observed in a rather large $H$ window sticked to $H_{\rm R}$ \cite{Levy2005}.
By a careful study of the resistivity under pressure and magnetic field, we have recently demonstrated that the SC properties can be well understood via a McMillan type formula, $T_{\rm sc} = T_0\exp(-m^{\ast}/m^{\ast\ast})$, where the effective mass $m^{\ast}$ is the sum of a renormalized band mass $m_{\rm B}$ and of a correlated mass $m^{\ast\ast}$ source of the SC pairing \cite{miyake2009}.
Field enhancement of $m^{\ast\ast}(H)$ will occur in the vicinity of $H_{\rm R}$, while $m_{\rm B}$ appears as field and pressure invariant.
Under pressure, $m^{\ast\ast}(0)$ decreases thus $T_{\rm sc}(m^{\ast\ast}(0))$ and the upper critical field ($H_{c2}(0, m^{\ast\ast}(0))$ decrease slowly.
The RSC is strongly affected by the concomitant decrease of $T_{\rm sc}(m^{\ast\ast}(H))$ and of $H_{\rm c2}\sim (m^{\ast}_H)^2T_{\rm sc}^2(m^{\ast}_H)$.
As predicted by our simple model, the RSC collapses at a pressure $P_{\rm RSC}\sim$~1.5~GPa far lower than the pressure $P_{\rm sc}$ where the low field SC will disappear, i.e. $m^{\ast\ast}(0)$ will collapse.
The low field SC seems to disappear only near 4~GPa (Fig.~\ref{URhGe}(b)) \cite{miyake2009}. 

\begin{figure}[h]
\begin{center}
\includegraphics[width=40pc]{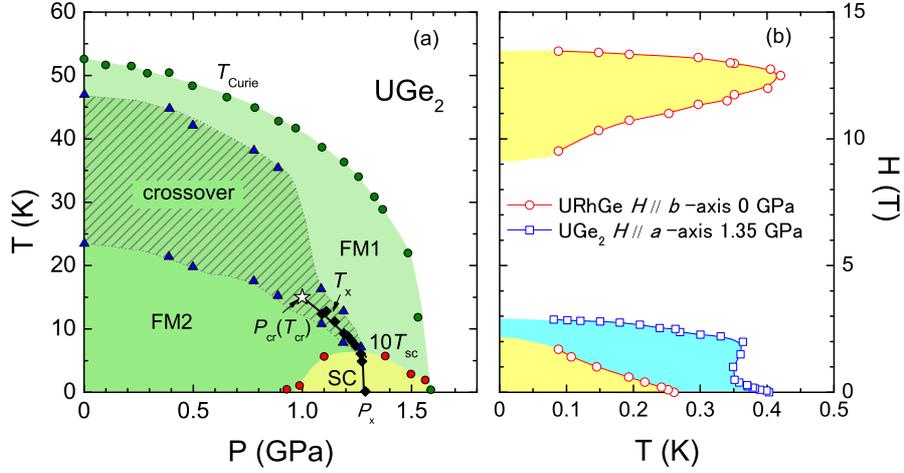}
\caption{\label{UGe2}
(a) ($P, T$) phase diagram of UGe$_{2}$ at $H = 0$~T.  The full diamond represents the $T_x (P)$ line, the star indicates a critical end point at $P_{\rm cr}$  = 1 GPa and $T_{\rm cr}$ = 15~K.
The crossover regime below $P_x$ is delimited by the width of the resistivity and thermal expansion bumps created by the proximity of the critical end point.
(b) Superconducting phase diagrams of URhGe at ambient pressure (open circle) and UGe$_2$ at 1.35~GPa (open square) after \cite{sheikin2001rcs}. }  
\end{center}
\end{figure}

In UGe$_2$, the SC dome is not linked to the proximity of the FM quantum singularity where FM will disappear but to the first order singularity at $P_x\sim$ 1.3~GPa where the system switches from a low pressure high sublattice magnetization FM2 phase ($M_0 = 1.5~\mu_{\rm B}$) to a high pressure low sublattice magnetization FM1 phase ($M_0 = 1~\mu_{\rm B}$). 
As shown in Fig.\ref{UGe2}(a), the new features are revealed that i) the first order line ($T_x, P_x$) between FM1 and FM2 terminates at a critical end point ($T_{\rm cr}$ =15~K, $P_{\rm cr}$=1~GPa) and ii) SC appears just above $P_{\rm cr}$.
The crossover from FM1 to FM2 due to a continuous mixing of both phase below $P_{\rm cr}$ was drawn from the broadening of the thermal expansion and the resistivity.
At ambient pressure, this definition corresponds to the upper limit of 10\% of emerging FM2 content in FM1 and the lower limit to 10\% remaining FM1 phase in FM2 as determined by combined specific heat and thermal expansion measurements \cite{Hardy}.
At a pressure $P_x + \epsilon\sim$1.35~GPa applying a magnetic field along the easy axis leads to switch from the initial FM1 to the FM2 state via a metamagnetic transition; the consequence for SC shown in Fig.\ref{UGe2}(b) is that the system reenters in the SC+FM2 phase, which has quite different parameters from SC+FM1; i.e. $m_{\rm B}$ and $m^{\ast\ast}$ (and thus $T_{\rm sc}$ and $H_{c2}$) \cite{sheikin2001rcs}. 
In contrast to URhGe, as $M_0$ is not preserved a drastic change of the Fermi surface occurs \cite{terashima2001eqp}. 
The concomitant variation of $m_{\rm B}$ and of $m^{\ast\ast}$ are marked by the fact above $P_x$ that its Sommerfeld coefficient is almost pressure invariant while $T_{\rm sc}(P)$ in the FM1 phase changes drastically \cite{Tateiwa2004}.
This simultaneous ($P, H$) variation of $m_{\rm B}$ and $m^{\ast\ast}$ was predicted in the theoretical model of Ref.~\cite{Sandeman2003} with a double peaked structure in the density of states.

These two examples of FM-SC corresponds to the case where $T_{\rm sc} \ll T_{\rm Curie}$.
With the recent case of UCoGe having $T_{\rm Cuire}\sim$~3~K and $T_{\rm sc}\sim$~0.6~K \cite{Huy2007}, the hope is to get new coexisting states when $T_{\rm Curie}$ and $T_{\rm sc}$ will cross under pressure as the possibility of a cascade from paramagnetic, superconducting and mixed SC+FM states \cite{mineev}.

\begin{figure}[h]
\begin{center}
\includegraphics[width=22pc]{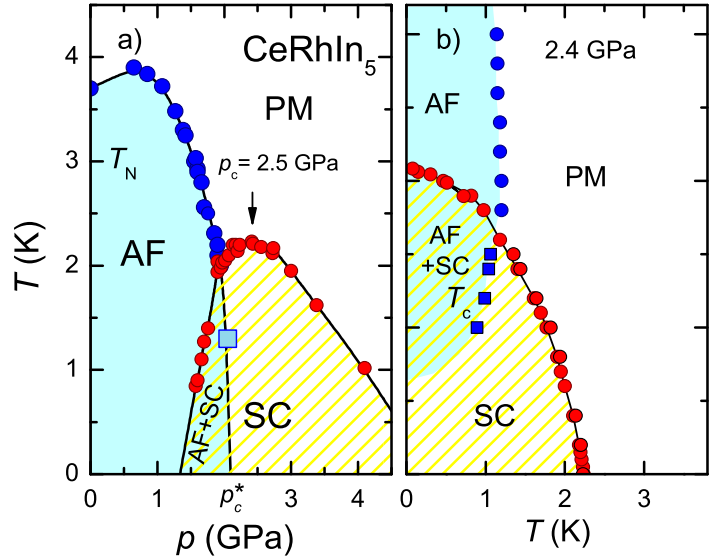}
\caption{\label{CeRhIn5}(a) ($P, T$) phase diagram of CeRhIn$_{5}$ at $H = 0$~T from ac calorimetry (circles) and from recent NQR experiments (square) \cite{Knebel2006, Yashima2007, Yashima2009}.  (b) ($T, H$) phase diagram at a pressure close to $P_c$ of $P =$ 2.4~GPa.}  
\end{center}
\end{figure}

As a guide line, it is interesting to look to careful studies realized under $(P, H)$ on CeRhIn$_5$ \cite{Knebel2004, Park2006, Knebel2008}.    
As shown in Fig.\ref{CeRhIn5}(a), antiferromagnetism (AF) and SC are robust at low and high pressure, respectively.
A coexisting SC+AF phase at zero field occurs only between $P\sim$~1.7~GPa and $P_c^{\ast}\sim$ 2~GPa, where $T_{\rm N}$ and $T_{\rm sc}$ merge into one point \cite{Knebel2008}.
The simple idea is a competition or interplay between an AF pseudogap $\Delta_{\rm AF}$ and the SC gap $\Delta_{\rm sc}$ assuming that both the AF and the SC phase will have the same (1/2, 1/2, 1/2) hot spot in the coexisting domain which seems to be shown by recent NQR results.\cite{Yashima2007, Yashima2009}.
Between $P^{\ast}_{c}<P<P_{c}\sim$~2.5~GPa, the new event is the field reentrance of AF under magnetic field and the associated creation of vortices \cite{Park2006, Knebel2008}, $T_{\rm N}(H)$ can exceed $T_{\rm sc}(H)$ as the critical magnetic field $H_{M}\sim$ 40~T at $T\to$~0~K is much larger than the value of the upper critical field $H_{c2}(0)\sim 10$~T, so far $T_{\rm N}$ does not collapse, i.e. $P$ is lower than $P_c$.
As indicated by the arrows in Fig.\ref{CeRhIn5}(b), the reentrant AF sticked to $H_{c2}(0)$ will collapse at $P_c$.
Up to now, attempts to fit the upper critical field curves $H_{c2}(T)$ via the orbital and Pauli competition plus additional possibility of strong coupling constant ($\lambda = m^{\ast}/m_{\rm B}-1$) have failed \cite{Knebel2008}; as discussed previously on FM-SC, for SC of CeRhIn$_5$ there is clearly the necessity to incorporate $H$ dependence of $m^{\ast}$ which is not only pressure dependent. 

It is worthwhile to mention that  CeCoIn$_5$ at $P$ = 0 may be just above $P^{\ast}_c$, and $H_{\rm M}$ must be lower than $H_{c2}(0)$; then the SC gap will allow to stick AF to $H_{c2}(0)$, so far $\Delta_{\rm sc}(H)$ is below a critical value $\Delta_c$ to allow the establishment of itinerant AF (see \cite{Knebel2008}).
Taking into account the strong coupling correlation, with this picture we arrive to the conclusion that increasing the pressure and thus decreasing the strength of $\lambda$, the AF phase will expand in a larger $H(P)/H_{c2}(P)$ domain than in zero pressure in good agreement with the experiment \cite{Knebel2008}.
Furthermore, when the SC gap at zero field will go down to $\Delta_c$, reentrant magnetism must disappear.

The comparison of the three cases where reentrant phenomena exists show clearly the importance to take into account the $P$ and also $H$ dependence of the effective mass. 
For FM-SC of URhGe the simplicity is that only the field dependence of the correlated mass $m^{\ast\ast}$ must be considered; in UGe$_2$ the $P$ and $H$ variation affect both contributions of the effective mass. 
With the case of CeRhIn$_5$ and CeCoIn$_5$, the interplay between an AF pseudogap and the SC gap appears the appealing key parameter.
Next interesting example to reveal the relation between magnetism and SC is a careful study of the FM superconductor UCoGe in the critical regime near 1~GPa where $T_{\rm sc} = T_{\rm Curie}$  \cite{Hassinger2008}.

\end{document}